# Graphene-based absorber exploiting guided mode resonances in one-dimensional gratings

M. Grande [a], M. A. Vincenti [b], T. Stomeo [c], G. V. Bianco [d], D. de Ceglia [b], N. Aközbek [e],
V. Petruzzelli [a], G. Bruno [d], M. De Vittorio [c,f], M. Scalora [g], A. D'Orazio [a]

[a] *Dipartimento di Ingegneria Elettrica e dell'Informazione, Politecnico di Bari,*
*Via Re David 200, 70125 Bari, Italy*
[b] *National Research Council, Charles M. Bowden Research Center, RDECOM,*
*Redstone Arsenal, Alabama 35898-5000 – USA*
[c] *Center for Bio-Molecular Nanotechnologies, Istituto Italiano di Tecnologia (IIT),*
*Via Barsanti, 73010 Arnesano (Lecce), Italy*
[d] *Institute of Inorganic Methodologies and of Plasmas, IMIP-CNR,*
*via Orabona 4, 70126 Bari, Italy*
[e] *AEgis Technologies Inc, 410 Jan Davis Dr, Huntsville - AL, 35806*
[f] *National Nanotechnology Laboratory (NNL), CNR-Istituto Nanoscienze, Dip. Ingegneria dell'Innovazione,*
*Università Del Salento, Via Arnesano, 73100 Lecce, Italy*
[g] *Charles M. Bowden Research Center, RDECOM,*
*Redstone Arsenal, Alabama 35898-5000 – USA*

**Abstract:** A one-dimensional dielectric grating, based on a simple geometry, is proposed and investigated to enhance light absorption in a monolayer graphene exploiting guided mode resonances. Numerical findings reveal that the optimized configuration is able to absorb up to 60% of the impinging light at normal incidence for both TE and TM polarizations resulting in a theoretical enhancement factor of about 26 with respect to the monolayer graphene absorption ($\approx$2.3%). Experimental results confirm this behaviour showing CVD graphene absorbance peaks up to about 40% over narrow bands of few nanometers. The simple and flexible design paves the way for the realization of innovative, scalable and easy-to-fabricate graphene-based optical absorbers.

## 1. Introduction

Graphene is a single atomic layer of graphite that consists of very tightly bonded carbon atoms organised into a hexagonal lattice [1]. Graphene shows an $sp^2$ configuration that leads to a total thickness of about 0.34 nm. This two-dimensional nature is responsible of the very exceptional electrical, mechanical and optical properties shown by this material.

In particular, it has been theoretically and experimentally demonstrated that the absorption of a monolayer graphene does not depend on the material parameters but only on the fundamental constants since it is equal to $\pi\alpha$ (defined by the fine structure constant $\alpha = e^2/\hbar c$) that corresponds to about 2.3% over the visible (VIS) range [2]. Moreover, the absorption of multiple graphene layers is proportional to the number N of added layers [2]. This important property has been exploited in different configurations in order to realize efficient graphene-based photo-detectors [3-4] and modulators [5-6]. At the same time, even if this constant value is very high when compared with other bulk materials, the absorption of monolayer graphene can be boosted and enhanced in different spectral ranges exploiting different technologies and approaches in both linear and nonlinear regimes. In particular, over the last few years, several solutions have been proposed by incorporating the monolayer graphene in configurations operating in the VIS and near-infrared ranges (NIR), that exploit attenuated total reflectance (ATR) [7] or resonant configurations such as one-dimensional (1D) periodic structures [8], two-dimensional photonic crystal cavities [9] and multilayer dielectric Bragg mirrors [10]. In this framework, in our previous work [11], we have shown how it is possible to achieve near perfect-absorption in a one-dimensional Photonic Crystal (PhC) that incorporates a graphene monolayer in the defect. Finally, enhanced absorption in monolayer graphene can also be achieved in the in mid-IR and terahertz (THz) ranges where this two-dimensional material shows a plasmonic behavior [12-16].

In this paper, we propose and investigate a one-dimensional dielectric grating that exploits guided mode resonances (GMRs) [17] to enhance light absorption in the monolayer graphene. Guided mode resonances define optical modes with complex wavenumber, typically leaky modes, which are strongly confined in the 1D grating. Forced excitation [18] of these lattice modes may be triggered by phase-matching to incident plane waves, thus, the interaction between these discrete modes and the out-of-plane radiation continuum gives rise to narrowband and asymmetric spectral features, also known as Fano resonances [19-20]. In this scenario, we will analyze absorption enhancement when the monolayer graphene is inserted in a lossless dielectric grating. In particular, we will numerically investigate the dependence of the absorption on the geometrical parameters with a plane wave excitation at normal incidence. Finally, we will detail the fabrication process and the experimental results related to the optical characterization of the device.

## 2. Numerical results

Figure 1 shows the sketch of the proposed 1D dielectric-grating-based absorber made of polymethyl-methacrylate (PMMA) stripes deposited on a tantalum pentaoxide ($Ta_2O_5$) slab that is supported by a silicon dioxide (SiO$_2$) substrate. The monolayer graphene is sandwiched between the polymeric layer and the Ta$_2$O$_5$ slab forcing it to interact with the guided mode resonances. Finally, the Ta$_2$O$_5$ slab thickness $t_{Ta2O5}$, the periodicity $p$, the PMMA width $w_{PMMA}$ and the PMMA thickness $t_{PMMA}$ are set initially equal to 100 nm, 470 nm, 235 nm ($w_{PMMA}$=0.6$p$) and 650 nm, respectively.

The one-dimensional grating has been simulated in the visible-near infrared (VIS-NIR) range and the dispersion for the different dielectric media has been experimentally measured by means of ellipsometric technique. It is worth stressing that these values almost coincide with the models based on the Sellmeier equation retrieved by the data reported in [21]. Furthermore, we found negligible losses for the Ta$_2$O$_5$ slab and PMMA layer (i.e., the extinction coefficients are equal to 0), hence the device without the graphene will be considered lossless hereinafter. Finally, the monolayer graphene has been modeled using the experimental fit reported in [22]. This model does not take into account the doping effect. In this respect, it is possible to refine the model following the Kubo formulation as proposed in [23] that considers the chemical potential, the temperature and the scattering time. However, in our range of interest the variation of the complex refractive index model is negligible (few percentage points) when the doping, i.e., the chemical potential, is varied. On the contrary, the model based on the Kubo formulation is essential for describing the graphene optical properties in the infrared and terahertz regimes. The spectral response of the configuration has been investigated for both TE and TM polarized incident plane waves. Figure 2 compares reflectance, transmittance and absorbance spectra without and with the inclusion of the monolayer graphene, respectively.

In particular, the device displays two asymmetric guided mode resonances located at about 743.9 nm and 712.6 nm for TE and TM polarization, respectively, with a full-width at half-maximum of few nanometers. The asymmetry is evident for the transmittance and reflectance curves while the absorption shows an almost symmetric response. The spectral position virtually satisfies the phase matching condition [18]:

$$n_{eff} = n_0 \sin\theta_{inc} - \frac{m\lambda_0}{p} \qquad (1)$$

where $n_{eff}$ corresponds to the effective refractive index of the mode in the slab, n$_0$ is the refractive index of cover medium, $\theta_{inc}$ is the angle of incidence of the impinging source, $m$ is the diffraction order, $\lambda_0$ is the free space wavelength and $p$ is the period of the grating.

At normal incidence (i.e. $\theta_{inc} = 0$) in air ($n_0 = 1$) and for the first diffraction order ($m = -1$), Equation (1) reduces to $n_{eff} = \lambda_0 / p$ and, hence, the free space wavelength at which the phase matching condition occurs is equal to $\lambda_0 = n_{eff} \cdot p$.

For the configuration reported in Figure 2, the effective refractive index, for TE and TM polarization, is equal to about 1.582 and 1.511, respectively, leading to a guided mode resonance located at about 743.5 nm and 710.2 nm. Therefore, the wavelength shift between the two polarizations (about 30 nm) is due to the different effective refractive index of the two modes. Further, the plots in Figure 2 clearly prove that the introduction of the monolayer graphene enhances the absorption of the device from 2.3% up to about 50% for both polarizations where this boost goes mainly to the detriment of reflectance more than the transmittance.

Then, the effects of the technological tolerances on the device performance have been investigated by varying the $Ta_2O_5$ thickness. Figures 3(a) and 3(c) depict the absorption maps of the 1D dielectric grating for TE and TM polarizations when the $Ta_2O_5$ thickness $t_{Ta2O5}$ is varied in the range 0-150 nm and the periodicity, the fill-factor (i.e. the PMMA width) and the PMMA thickness are left unchanged.

The maps reveal that the wavelengths associated with the guided mode resonance shift almost linearly with the $Ta_2O_5$ slab thickness (Figures 3(a)-(c)). Moreover, the absorption for TM polarization shows a narrower bandwidth with respect to its TE counterpart, consistent with the spectra reported in Figure 2. At the same time, absorption abruptly increases when the thickness overcomes a threshold for both polarizations due to the different effective refractive index (Figures 3(b)-(d)). After the threshold related to the mode cutoff, the absorption is rather flat for TM polarization while it slightly varies for TE polarization. Therefore, for a $Ta_2O_5$ slab thickness $t_{Ta2O5}$= 100 nm one can excite both guided mode resonances.

A similar approach can be employed for the analysis of the optical response of the device when the PMMA thickness is varied. Figure 4 depicts the behavior of the device in this circumstance. Also in this case, the maps shown in Figures 4(a)-(b) reveal that the absorption remains almost unaltered with small variations. At the same time, the wavelength associated with the guided mode resonance is unaffected. Consequently, the device is not very sensitive to the PMMA layer thickness variations over a very large range (about 500 nm). This behavior is in agreement with the results reported in [15] explaining that it is sufficient an initial thickness to excite the guided mode resonance; thus, when the thickness is increased, the modal configuration is unaffected and, hence, the optical response does not change in a noticeable way. Similar considerations can be done for the reflectance spectra (Figures 4(c)-(d)). In conclusion, these maps indicate that the maximum attainable absorption for this configuration is about 60% corresponding to an enhancement factor of about 26 with respect to the monolayer graphene absorption ($\approx 2.3\%$).

## 3. Experimental results

In order to verify the numerical findings, the device under examination has been fabricated. In particular, a 100 nm-thick $Ta_2O_5$ slab was grown on a $SiO_2$ substrate by means of a RF sputtering system. The sample has been treated by means of oxygen plasma in order to increase the wettability and improve the adhesion properties. Then, a monolayer graphene, grown via Chemical Vapour Deposition (CVD) technique, was manually transferred onto the $Ta_2O_5$ slab. The 1D grating was realized in two steps: firstly, a 650 nm-thick PMMA layer was spin-coated onto the sample, and then the PMMA layer was exposed by means of an electron beam lithography system (Raith150) operating at 20 kV. Finally, the sample was developed in a methyl isobutyl ketone – isopropyl alcohol (MIBK-IPA) (MIBK-IPA) mixture and rinsed in an IPA bath. It is worth pointing out that we set the PMMA thickness equal to about 650 nm since this thickness corresponds to a maximum in the reflectance maps for both polarizations as shown in Figures 4(c)-(d). We retained this necessary in order to verify the presence of the residual reflectance peaks.

Figure 5(a) shows the Scanning Electron Microscope (SEM) image of the final device revealing the periodic PMMA stripes. The quality of the monolayer graphene was verified, after the fabrication process, by means of a Horiba Jobin-Yvon LabRAM HR-VIS micro-Raman spectrometer equipped with a 532 nm laser source. The Raman spectrum of the processed monolayer graphene is reported in Figure 5(b) proving the virtual absence of defects (i.e. D peak at about 1350 $cm^{-1}$), hence the preservation of the graphene quality.

The fabricated device was optically characterized at normal incidence by means of an optical setup constituted by a white-light lamp, filtered in the 600 nm–900 nm range, focused on the sample by means of a low numerical aperture, infinity-corrected microscope objective (5X, NA=0.15) [24]. The reflected light was collected by an aspherical fiber lens collimator and filtered by a linear polarizer. An opaque metallic stage was used to support the sample and avoid collecting unwanted light. The filtered light was sent to an optical spectrometer (HR4000 from

Ocean Optics) through a multimode optical fiber. The reflectance spectrum normalization was carried out by using a flat silicon surface as spectrum reference. At the same time, the transmittance spectra were collected through a perforated stage allowing the light to reach another aspherical fiber lens collimator. In this case, the $Ta_2O_5$ substrate sample was used as reference.

Figure 6 shows the comparison between the theoretical and experimental results revealing an excellent agreement when the periodicity *p* is slightly tuned by few nanometers. The experimental wavelengths for the guided mode resonances are located at about 737 nm and 695 nm, respectively and the TE polarization related peak shows a broader bandwidth (about 7 nm) with respect to the TM counterpart (about 3 nm). Furthermore, the experimental absorption is equal to 43% (34%) for TE (TM) polarization corresponding to an enhancement factor of about 19 (15) with respect to the absorption of a monolayer graphene (equal to 2.3%).

## 4. Conclusion

We have detailed the design, fabrication and optical characterization of a one-dimensional dielectric grating based absorber that incorporates a monolayer graphene and exploits polymeric stripes in order to excite guided mode resonances in the VIS-NIR range. The numerical findings reveal that a single layer of graphene suffices to absorb about 60% of the impinging light at normal incidence for both polarizations over a narrow bandwidth of few nanometers. Further, the device is less sensitive to the $Ta_2O_5$ slab thickness variation when this thickness exceeds a very sharp knee that defines a threshold due to the mode cutoff. Additionally, the optical response of the device is not affected even when the PMMA thickness is varied showing an almost flat absorption over a very large range with negligible variations.

The optical characterization revealed an excellent agreement between the experimental data and the theoretical results confirming the ability of this graphene-based absorber to achieve an enhancement factor of about 19 (15) for TE (TM) polarization. It is worth stressing that we have presented a 1D grating but the same idea can be extended to 2D arrays of square dielectric patches in a similar fashion as reported in [24]: in this way it could be possible to realize devices that are less sensitive to the incident light polarization.

In conclusion, we have presented an approach to achieve enhanced absorption in graphene exploiting guided mode resonances. A fair comparison between our previous work reported in

[11] and the proposed device allows us to draw some conclusions: the defective photonic-crystal configuration can achieve perfect absorption while the guided-mode resonance approach shows a maximum absorption of about 60%. However, the latter is based on a very simple geometry and shows low sensitivity to the geometrical parameters such as the PMMA thickness, hence significant robustness in terms of fabrication tolerances. Additionally, the proposed device requires polymeric stripes that could be easily realized by low cost and scalable fabrication technologies such as the nano-imprinting lithography. The 1D-PhC also requires low intrinsic dielectric losses due to the high number of layers in the dielectric stack while, here, losses depend only on two dielectric slabs. Therefore, this device could be efficiently exploited as building blocks for innovative optical absorbers or photo-detectors in combination with active materials (e.g. silicon photonics based devices). Finally, the proposed solution appears also interesting in order to enhance the exceptional nonlinear third harmonic and the saturable responses of the monolayer graphene [25-26] in a similar fashion reported in Ref. [11,27].

**Acknowledgements**

This research was performed while the authors M. A. Vincenti and D. de Ceglia held National Research Council Research Associateship awards at the U. S. Army Aviation and Missile Research Development and Engineering Center. M. Grande thanks the U.S. Army International Technology Center Atlantic for financial support (W911NF-13-1-0434).

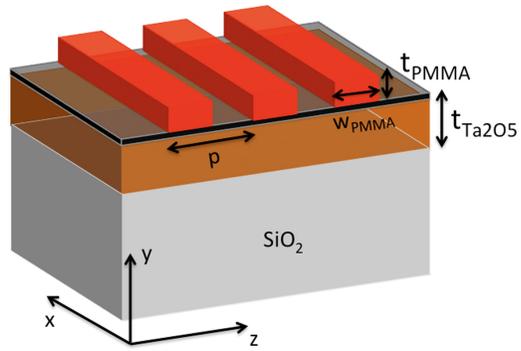

Fig. 1. Sketch of the 1D grating: PMMA stripes (red) on Ta$_2$O$_5$ slab (orange) grown on silicon dioxide substrate (grey). The black thin layer indicates the monolayer graphene.

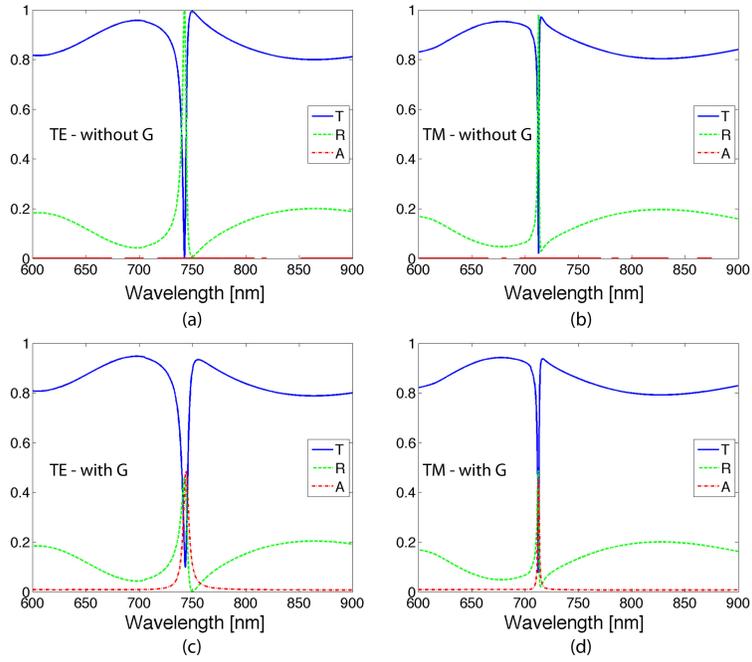

Fig. 2. Transmittance (blue solid line, T), Reflectance (green dashed curve, R) and Absorbance (red dash-dot curve, A) without (top panels) and with (bottom panels) the monolayer graphene for the TE (a-c) and TM (b-d) polarizations when $t_{Ta2O5}$, $p$, $w_{PMMA}$ and $t_{PMMA}$ are equal to 100 nm, 470 nm, 235 nm (0.6p) and 650 nm, respectively.

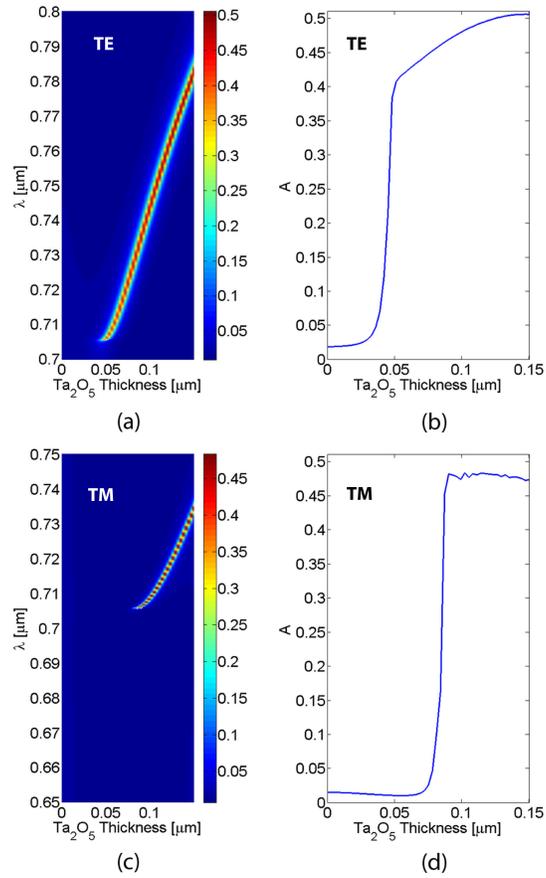

Fig. 3. Absorption maps for the (a) TE and (c) TM polarization, respectively, when the $Ta_2O_5$ thickness is varied. Maximum achievable absorption versus the $Ta_2O_5$ thickness for the (b) TE and (d) TM polarization, respectively.

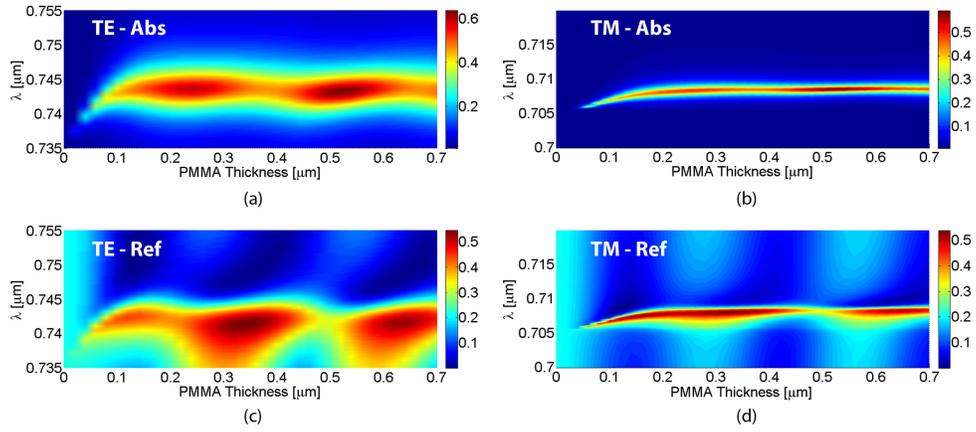

Fig. 4. Absorbance maps for the 1D dielectric grating when the PMMA thickness $t_{PMMA}$ is varied between 0 nm and 700 nm for the (a) TE and (b) TM polarization, respectively. Panels (c) and (d) refer to the corresponding reflectance maps for the TE and TM polarization, respectively.

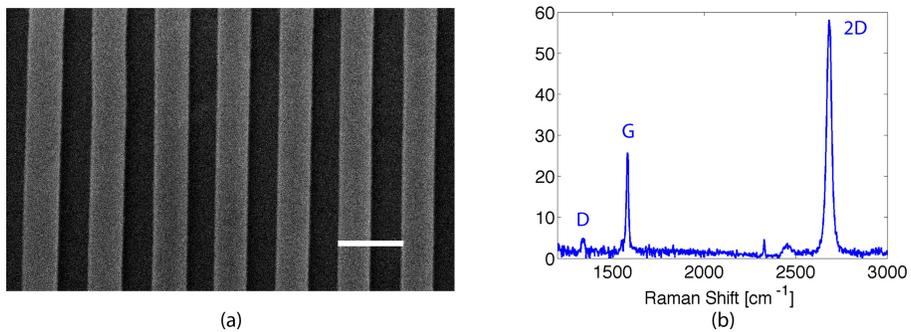

Fig. 5. (a) Scanning Electron Microscope (SEM) micrograph of the fabricated device; the white scalebar refers to 470 nm. (b) Raman spectrum of the monolayer graphene after the fabrication process.

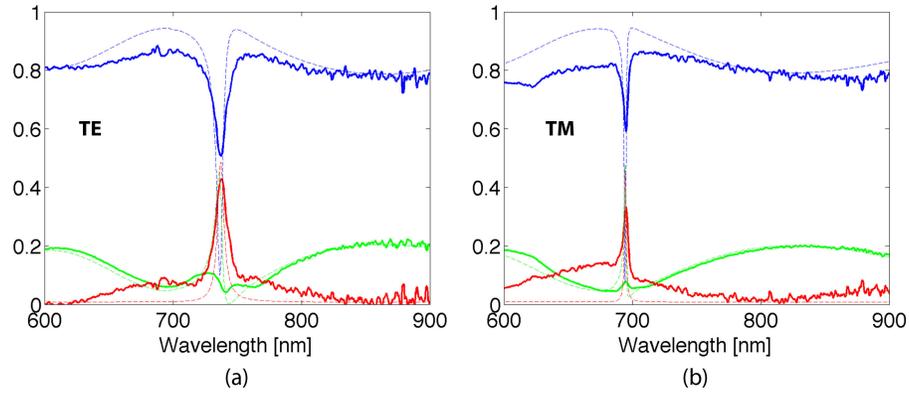

Fig. 6. Theoretical (dashed lines) and experimental (solid lines) Transmittance (blue curves), Theoretical reflectance (green curves) and Absorbance (red curves) spectra for the TE (a) and TM (b) polarizations when when $t_{Ta2O5}$, $w_{PMMA}$ and $t_{PMMA}$ are equal to 100, 235 nm (0.6p) and 650 nm, respectively.